\documentclass[aps,prd,nofootinbib,twocolumn,superscriptaddress,preprintnumbers]{revtex4}

\usepackage{amssymb}
\usepackage{amsmath}
\usepackage{epsfig}
\usepackage{hyperref}
\usepackage{breakurl}

\makeatletter
\def\simgt{\mathrel{\lower2.5pt\vbox{\lineskip=0pt\baselineskip=0pt
           \hbox{$>$}\hbox{$\sim$}}}}
\def\simlt{\mathrel{\lower2.5pt\vbox{\lineskip=0pt\baselineskip=0pt
           \hbox{$<$}\hbox{$\sim$}}}}
\makeatother

\newcommand{\be}{\begin{equation}}
\newcommand{\ee}{\end{equation}}
\newcommand{\bea}{\begin{eqnarray}}
\newcommand{\eea}{\end{eqnarray}}
\newcommand{\Eq}[1]{Eq.~(\ref{#1})}
\newcommand{\Eqs}[2]{Eqs.~(\ref{#1}) and (\ref{#2})}
\newcommand{\Sec}[1]{Sec.~\ref{#1}}

\newcommand{\Fig}[1]{Fig.~\ref{#1}}
\newcommand{\Figs}[2]{Figs.~\ref{#1} and \ref{#2}}

\newcommand{\Feff}{F_{\rm tot}}
\newcommand{\MPl}{M_{\rm Pl}}

\begin{document}

\preprint{UCB-PTH-10/09}
\preprint{MIT-CTP 4144}

\title{A Definitive Signal of Multiple Supersymmetry Breaking}

\author{Clifford Cheung}
\affiliation{Berkeley Center for Theoretical Physics, 
  University of California, Berkeley, CA 94720, USA}
\affiliation{Theoretical Physics Group, 
  Lawrence Berkeley National Laboratory, Berkeley, CA 94720, USA}

\author{Jeremy Mardon}
\affiliation{Berkeley Center for Theoretical Physics, 
  University of California, Berkeley, CA 94720, USA}
\affiliation{Theoretical Physics Group, 
  Lawrence Berkeley National Laboratory, Berkeley, CA 94720, USA}

\author{Yasunori Nomura}
\affiliation{Berkeley Center for Theoretical Physics, 
  University of California, Berkeley, CA 94720, USA}
\affiliation{Theoretical Physics Group, 
  Lawrence Berkeley National Laboratory, Berkeley, CA 94720, USA}
\affiliation{Institute for the Physics and Mathematics of the Universe, 
  University of Tokyo, Kashiwa 277-8568, Japan}

\author{Jesse Thaler}
\affiliation{Center for Theoretical Physics, 
  Massachusetts Institute of Technology, Cambridge, MA 02139, USA}

\begin{abstract}
If the lightest observable-sector supersymmetric particle (LOSP) is charged and long-lived, 
then it may be possible to indirectly measure the Planck mass at the LHC 
and provide a spectacular confirmation of supergravity as a symmetry of 
nature.  Unfortunately, this proposal is only feasible if the gravitino 
is heavy enough to be measured at colliders, and this condition is 
in direct conflict with constraints from big bang nucleosynthesis (BBN). 
In this work, we show that the BBN bound can be naturally evaded in the presence of
multiple sectors which independently break supersymmetry, since there is a new decay channel of the LOSP to a goldstino.  Certain regions 
of parameter space allow for a direct measurement of LOSP decays into 
both the goldstino and the gravitino at the LHC.  If the goldstino/gravitino 
mass ratio is measured to be 2, as suggested by theory, then this would 
provide dramatic verification of the existence of multiple supersymmetry 
breaking and sequestering.  A variety of consistent cosmological scenarios 
are obtained within this framework.  In particular, if an $R$ symmetry 
is imposed, then the gauge--gaugino--goldstino interaction vertices can 
be forbidden.  In this case, there is no bound on the reheating temperature 
from goldstino overproduction, and thermal leptogenesis can be accommodated 
consistently with gravitino dark matter.
\end{abstract}

\maketitle

\section{Introduction}

At first glance, particle colliders do not appear particularly well-suited 
to the task of probing the fundamental structure of gravity.  Indeed, 
given the intrinsic energy limitations of present day and future machines, 
any direct experimental handle on Planck scale physics continues to be 
a rather remote possibility.  Nevertheless, weak scale supersymmetry (SUSY) 
may provide a unique window into genuinely gravitational physics because 
SUSY is a symmetry of spacetime.

In particular, if SUSY is realized as a local symmetry, namely supergravity 
(SUGRA), then there necessarily exists a spin-$3/2$ superpartner of the 
graviton: the gravitino.  The gravitino has a mass, $F/\sqrt{3}\MPl$, 
and couplings to observable sector fields, $\sim 1/F$, which obey 
a fixed relationship determined by the Planck scale, $\MPl$.  Because 
the existence of the gravitino is mandatory, it is an attractive 
possibility that this state comprises the dark matter of the universe. 
In this case, the lightest observable-sector supersymmetric particle 
(LOSP) may be charged, and precision studies of the LOSP decay into 
the gravitino can provide a robust test of the expected gravitino 
mass/interaction relation and therefore an indirect measurement of 
$\MPl$~\cite{Buchmuller:2004rq,Hamaguchi:2006vu}.  This would offer 
compelling evidence for the validity of SUGRA as well as a genuine 
probe of gravitational physics at particle colliders.

Unfortunately, this most spectacular signal is in direct conflict 
with big bang nucleosynthesis (BBN).  At colliders, $\MPl$ can only 
be measured if the gravitino is sufficiently heavy: $m_{3/2} \simgt 
O(0.1) m_{\rm LOSP}$.  However, unless the gravitino is sufficiently 
light, $m_{3/2} \simlt O(1~\mbox{--}~10~{\rm GeV})$, then late-time 
charged LOSP decays destroy the successful predictions for the 
abundance of light elements~\cite{Khlopov:1984pf,Kawasaki:2008qe}. 
Thus, if the gravitino is to be heavy enough for a successful 
collider measurement, one must resort to a rather non-standard 
cosmology in which the thermal history is modified below a 
temperature of $O(0.1) m_{\rm LOSP}$.  Additionally, thermal 
leptogenesis~\cite{Fukugita:1986hr}---arguably the simplest mechanism 
for baryogenesis---does not work for the small values of $m_{3/2}$ 
required by the BBN constraint unless the gravitino is extremely 
light ($m_{3/2} \simlt 10~{\rm eV}$).  This is a consequence 
of gravitino overproduction from the high temperature 
plasma~\cite{Moroi:1993mb,Rychkov:2007uq} and constraints 
from structure formation~\cite{Viel:2005qj}.

It is interesting to note, however, that the BBN bound actually has 
nothing to do with the gravitino itself---it has to do with the late 
decaying LOSP injecting energies during or after BBN.  Therefore, if 
there is an additional state to which the LOSP can decay more quickly, 
then the constraint from BBN may be avoided.  However, this new state 
often introduces its own cosmological problems, and is not necessarily 
theoretically motivated.

In this paper, we show that a promising new state does in fact exist: an 
uneaten goldstino.  This mode arises naturally in the general framework 
of Ref.~\cite{Cheung:2010mc}, where multiple sectors separately break SUSY, 
yielding a corresponding multiplicity of goldstini.  In this framework, 
the gravitino couplings are not modified, but the LOSP can decay faster 
to an uneaten goldstino, nullifying the usual BBN constraint.  Moreover, 
in the limit where the SUSY breaking sectors are mutually sequestered, 
the goldstini acquire a mass from SUGRA effects which is exactly twice 
the gravitino mass.  Intriguingly, this factor of $2$ is entirely fixed 
by the symmetries of SUGRA.

\begin{figure}
\begin{center}
\includegraphics[scale=1]{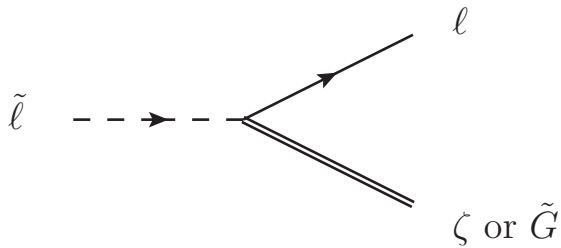}
\end{center}
\caption{Slepton LOSP decay into a goldstino or a gravitino.}
\label{fig:LOSP-decay}
\end{figure}
As we will see, the scenario outlined above leads to completely 
consistent cosmologies with a heavy gravitino $m_{3/2} \approx 
O(10~\mbox{--}~100~{\rm GeV})$.  This allows one to probe SUGRA via 
precision studies of charged LOSP decays at the LHC.  Furthermore, the 
LOSP will generically have non-negligible branching fractions into both 
the gravitino and a goldstino (see \Fig{fig:LOSP-decay}), allowing for 
a measurement of the relative factor of $2$ between their masses. 
Measuring $\MPl$ and observing this ``smoking gun'' factor of $2$ would 
reveal a number of exceedingly deep facts about our universe---not 
only that SUGRA is correct, but also that SUSY breaking is a generic 
phenomenon and sequestering is realized in nature.  These in turn 
suggest the existence of extra dimensions in which sequestering is naturally realized.

We find it remarkable that consistent cosmological histories happen to 
favor regions of parameter space in which dramatic LHC signatures are 
accessible.  In fact, by requiring thermal leptogenesis, the gravitino mass 
must be larger than $\approx 10~{\rm GeV}$, with the right abundance for 
gravitino dark matter obtained for reheating temperatures $T_R \approx 
(10^9~\mbox{--}~10^{10})~{\rm GeV}$.  This setup is possible if the 
goldstino interactions satisfy certain simple conditions, namely that 
they preserve an $R$ symmetry.  More generally, consistent cosmologies 
are obtained with $T_R$ as high as $\approx 10^7~{\rm GeV}$, in which 
case dark matter is dominantly the goldstino.  In both these cases, 
cosmology prefers the LOSP branching ratios into gravitinos and 
goldstinos to be not too dissimilar.

The remainder of the paper is organized as follows.  We review the 
collider tests of SUGRA and their tension with BBN in \Sec{sec:general}, 
and show how modified LOSP decays can simply relieve this tension.  In 
\Sec{sec:goldstini}, we summarize the framework of multiple SUSY 
breaking and the important properties of the resulting goldstini~\cite{Cheung:2010mc}. 
The collider phenomenology of gravitinos/goldstinos is discussed in 
\Sec{sec:collider}, and their cosmology is studied in \Sec{sec:cosmology}, 
where the relevant calculation of goldstino relic abundance is summarized 
in the Appendix.  The possibility of colored LOSPs is discussed in 
\Sec{sec:others}.  Finally, we conclude in \Sec{sec:concl}.

\section{BBN and the LHC}
\label{sec:general}

Testing the relationship between the mass and interaction strength of 
the gravitino requires precision collider measurements which are only 
feasible if the LOSP is charged.  Typically, the LOSP is taken to be 
a long-lived slepton (most commonly a stau) because this is favored in 
many SUSY breaking mediation schemes.  While other charged states are 
also possible, we will mostly focus on a slepton LOSP in this paper, 
leaving a discussion of other possibilities to \Sec{sec:others}.

At the LHC, quasi-stable charged sleptons may be copiously produced 
at the end of SUSY cascade decays.  Most of them will exit the 
interaction region appearing as ``heavy muons.''  The slepton mass 
can then be determined from a combination of time-of-flight and momentum 
information~\cite{Hinchliffe:1998ys}.  Through ionization energy 
loss, a fraction of the produced sleptons will also be trapped 
inside the main detector~\cite{Asai:2009ka} or in a separate stopper 
detector~\cite{Hamaguchi:2004df,Hamaguchi:2006vu}, where their 
decays can be precisely studied.

The decay rate of the slepton is measured by observing the lifetime 
of stopped sleptons.  In the conventional SUSY setup, a slepton decays 
to a lepton and a gravitino, with a width given by
\be
  \Gamma_{\tilde{\ell} \rightarrow \ell \tilde{G}} 
  \simeq \frac{m_{\tilde{\ell}}^5}{16\pi \Feff^2},
\label{eq:Gamma}
\ee
where $\Feff$ is the scale of SUSY breaking, and we have ignored the 
relatively unimportant phase space factor.  From the energy spectrum 
of the outgoing lepton and the measured slepton mass, one can also 
determine the gravitino mass, which is fixed by theory to be
\be
  m_{3/2} \simeq \frac{\Feff}{\sqrt{3} \MPl}.
\label{eq:m32}
\ee
The gravitino mass/interaction relation can then be 
tested~\cite{Buchmuller:2004rq} by combining the measured values 
of $\Gamma_{\tilde{\ell} \rightarrow \ell \tilde{G}}$ and $m_{3/2}$ 
to form the Planck scale $\MPl$
\be
  \MPl^2 \simeq \frac{m_{\tilde{\ell}}^5} 
    {48\pi \Gamma_{\tilde{\ell} \rightarrow \ell \tilde{G}} m_{3/2}^2},
\label{eq:measure-MPl}
\ee
and comparing it with the value obtained in long distance measurements 
of gravity.

However, an indirect measurement of $\MPl$ is only possible if the mass 
of the gravitino is sufficiently heavy that it can be experimentally 
determined.  In particular, for a given slepton decay, $m_{3/2}$ is 
reconstructed from the slepton mass and the energy of the outgoing 
lepton $E_\ell$ according to
\be
  m_{3/2} = \sqrt{m_{\tilde{\ell}}^2 + m_\ell^2 
    - 2 m_{\tilde{\ell}} E_\ell}.
\label{eq:m32-exp}
\ee
Thus, the error in $m_{3/2}$ is given by
\be
  \frac{\varDelta m_{3/2}}{m_{3/2}} \simeq 
    \frac{m_{\tilde{\ell}}^2}{2 m_{3/2}^2} ( \varDelta_m \oplus \varDelta_E),
\label{eq:m32-error}
\ee
where $\varDelta_m \equiv \varDelta m_{\tilde{\ell}}/m_{\tilde{\ell}}$ and 
$\varDelta_E \equiv \varDelta E_\ell/E_\ell$.  With sufficient statistics, 
we expect that $\varDelta m_{\tilde{\ell}}$ and $\varDelta_E$ can reach  
the level of $(0.1~\mbox{--}~1)\%$ at the LHC~\cite{Hinchliffe:1998ys,TDR}. 
This implies that the $\MPl$ measurement is possible for
\be
  m_{3/2} \simgt (0.05~\mbox{--}~0.2) m_{\tilde{\ell}},
\label{eq:m32-lower}
\ee
where we have required $\varDelta m_{3/2} \ll m_{3/2}$.  Of course, the 
precise numbers are subject to the level of experimental accuracy which 
may ultimately be achieved.

In standard SUSY, the mass and interaction strength of the gravitino 
obey a fixed relation, so any theory with a gravitino heavy enough to 
satisfy \Eq{eq:m32-lower} will have commensurately long-lived sleptons. 
As a consequence, this class of theories is in direct tension with 
BBN.  Specifically, sleptons produced in the early universe will decay 
during or after BBN and potentially alter the abundances of light 
elements.  As seen in Ref.~\cite{Kawasaki:2008qe}, the BBN constraint 
on late-decaying slepton LOSPs implies
\be
  m_{3/2} \simlt 
  \left\{ \begin{array}{ll}
    0.35~{\rm GeV} \left(\frac{m_{\tilde{\ell}}}{100~{\rm GeV}}\right)^{2.3} 
    & \mbox{ for }\, m_{\tilde{\ell}} \simlt 400~{\rm GeV},\\ \\
    20~{\rm GeV} \left(\frac{m_{\tilde{\ell}}}{1~{\rm TeV}}\right)^{1.0} 
    & \mbox{ for }\, m_{\tilde{\ell}} \simgt 400~{\rm GeV}.
  \end{array} \right.
\label{eq:m32-upper}
\ee
Here a typical primordial slepton yield of $Y_{\tilde{\ell}} \simeq 7 \times 
10^{-14} (m_{\tilde{\ell}}/100~{\rm GeV})$ was assumed, but the bound 
depends only weakly on this value.  Thus, one sees that the criterion for  
measuring $\MPl$ at colliders, \Eq{eq:m32-lower}, is in conflict 
with the BBN bound in conventional SUSY theories.  Note that the 
couplings of the gravitino are completely fixed by $m_{3/2}$, so there 
is no freedom to modify the slepton decay width to the gravitino.

Nevertheless, while the {\it partial width} of the slepton into the 
gravitino is fixed by SUGRA, it is of course possible to change the 
{\it total width} of the slepton.  By introducing a new light degree 
of freedom $\zeta$ and a new decay mode
\be
  \tilde{\ell} \rightarrow \ell \zeta,
\label{eq:LOSP-zeta}
\ee
the slepton can decay much more quickly and thus evade constraints from 
BBN.  In fact, a new field $\zeta$ arises quite naturally in the framework 
of multiple sector SUSY breaking, where $\zeta$ is identified as an 
{\it uneaten} goldstino.  We will review this framework in the next section.

In order to test the gravitino mass/interaction relation at the LHC, 
it is necessary that the slepton has a non-negligible branching 
fraction to gravitinos.  Assuming that there are $O(10^3~\mbox{--}~10^4)$ 
stopped LOSPs (which correspond to relatively light 
superpartners with $(100~\mbox{--}~1000)~{\rm fb}^{-1}$ of integrated 
luminosity~\cite{Hamaguchi:2006vu}) we need ${\rm Br}_{\tilde{\ell} 
\rightarrow \ell \tilde{G}} \simgt O(10^{-4}~\mbox{--}~10^{-3})$.

In summary, in order to measure $\MPl$ at colliders and simultaneously evade constraints from BBN, the following conditions must be satisfied:
\be
  \Gamma_{\tilde{\ell} \rightarrow \ell \zeta} \simgt 
  \left\{ \begin{array}{ll}
    9.1 \!\times\! 10^{-29}~{\rm GeV} 
      \left(\frac{m_{\tilde{\ell}}}{100~{\rm GeV}}\right)^{0.4} 
    & \!\! (m_{\tilde{\ell}} \simlt 400~{\rm GeV}), \\ \\
    2.8 \!\times\! 10^{-27}~{\rm GeV} 
      \left(\frac{m_{\tilde{\ell}}}{1~{\rm TeV}}\right)^{3.0} 
    & \!\! (m_{\tilde{\ell}} \simgt 400~{\rm GeV}),
  \end{array} \right.
\label{eq:BBN}
\ee
\begin{align}
  \frac{\Gamma_{\tilde{\ell} \rightarrow \ell \tilde{G}}} 
    {\Gamma_{\tilde{\ell} \rightarrow \ell \zeta}} 
  &\simgt O(10^{-4}~\mbox{--}~10^{-3}),
\label{eq:rate-obs}\\
  m_{3/2} &\simgt (0.05~\mbox{--}~0.2) m_{\tilde{\ell}},
\label{eq:mass-obs}
\end{align}
where the first condition has been translated from \Eq{eq:m32-upper} and 
so has a mild dependence on the primordial LOSP yield $Y_{\tilde{\ell}}$.

As shown in Ref.~\cite{Cheung:2010mc} and reviewed below, the mass of an uneaten 
goldstino is fixed by the symmetries of SUGRA to be $2m_{3/2}$. 
Consequently, if the gravitino mass is heavy enough to be determined 
at colliders, then so too is the mass of the goldstino.  Thus, we are 
presented with the intriguing prospect of measuring {\it both} decay 
channels to gravitino and goldstino, as well as the remarkable factor of $2$ in the 
mass relation.

\section{Review of Goldstini Framework}
\label{sec:goldstini}

In principle, any light mode $\zeta$ which couples with sufficient strength to 
the LOSP can nullify BBN constraints.  Here we will focus on the framework 
introduced in Ref.~\cite{Cheung:2010mc}, where $\zeta$ is an uneaten 
goldstino which arises in the context of multiple sector SUSY breaking. 
We find this a particularly attractive possibility both because it is 
well-motivated from top-down considerations and because it allows for 
a direct experimental probe of the fundamental properties of spacetime 
at colliders.  In what follows, we briefly review the case of two sectors 
which independently break SUSY, and refer the interested reader to 
Ref.~\cite{Cheung:2010mc} for a significantly more detailed treatment.

Consider two sectors which separately experience $F$-term SUSY breaking 
at the scales $F_1$ and $F_2$, yielding two corresponding goldstini fields, 
$\eta_1$ and $\eta_2$.  (We take $F_1 > F_2$ without loss of generality.) 
Because SUSY is a local symmetry, a diagonal combination of these goldstini 
is eaten by the gravitino via the super-Higgs mechanism, while the 
remaining orthogonal mode persists as a physical degree of freedom. 
We can go to the physical mass basis via the transformation
\bea
  \left( \begin{array}{c}
    \eta_1 \\ \eta_2 
  \end{array} \right)
&=&
  \left( \begin{array}{cc}
    \cos\theta & -\sin\theta \\
    \sin\theta &  \cos\theta 
  \end{array} \right) 
  \left( \begin{array}{c}
    \eta_{\rm long} \\ \zeta 
  \end{array} \right),
\label{eq:goldstinibasis}
\eea
where $\tan\theta = F_2/F_1$.  Here $\eta_{\rm long}$ is the longitudinal 
mode of the gravitino, while $\zeta$ is the uneaten goldstino which 
remains in the spectrum.  Since the overall scale of SUSY breaking 
is $\Feff = \sqrt{F_1^2+F_2^2}$, the gravitino mass is $m_{3/2} = 
\Feff/\sqrt{3} \MPl$.

The couplings of each goldstino to chiral and vector superfields of the 
supersymmetric standard model (SSM) are
\bea
  {\cal L}_{\phi} &=& \left( \frac{\tilde{m}_1^2}{F_1} \eta_1 
    + \frac{\tilde{m}_2^2}{F_2} \eta_2 \right) \psi \phi^\dagger 
    + {\rm h.c.},
\label{eq:goldstini-matter_1}\\
  {\cal L}_{\lambda} &=& -\frac{i}{\sqrt{2}} \left( \frac{M_1}{F_1} \eta_1 
    + \frac{M_2}{F_2} \eta_2 \right) \sigma^{\mu\nu} \lambda F_{\mu\nu} 
    + {\rm h.c.},
\label{eq:goldstini-gauge_1}
\eea
where $\phi$, $\psi$, and $\lambda$ represent SSM scalars, fermions, 
and gauginos, respectively.  The soft mass terms $\tilde{m}_{1,2}^2$ and 
$M_{1,2}$ are the contributions to the scalar squared masses and gaugino 
masses from sectors 1 and 2, respectively.  We primarily consider 
a regime in which the SUSY breaking scale of sector~1 is sufficiently 
larger than that of sector~2, so that $\Feff \approx F_1 \gg F_2$. 
In this limit, the couplings are
\begin{align}
  {\cal L}_{\phi} &\approx
    \left( \frac{\tilde{m}_1^2+\tilde{m}_2^2}{\Feff} \eta_{\rm long} 
    + \frac{\tilde{m}_2^2}{F_2} \zeta \right) \psi \phi^\dagger 
    + {\rm h.c.},
\label{eq:goldstini-matter_2}\\
  {\cal L}_{\lambda} &\approx 
    -\frac{i}{\sqrt{2}} \left(\frac{M_1+M_2}{\Feff} \eta_{\rm long} 
    + \frac{M_2}{F_2} \zeta \right) \sigma^{\mu\nu} \lambda F_{\mu\nu} 
    + {\rm h.c.}
\label{eq:goldstini-gauge_2}
\end{align}
As long as $\tilde{m}_2^2$ and $M_2$ are not too small, the SSM fields 
will couple more strongly to the uneaten goldstino $\zeta$ than to 
the longitudinal mode of the gravitino $\eta_{\rm long}$, allowing for 
substantial departures from usual SUGRA signatures.

In Ref.~\cite{Cheung:2010mc}, it was shown that $\zeta$ (and more 
generally, any additional uneaten goldstini) acquires a mass
\be
  m_\zeta = 2 m_{3/2} + \delta m,
\label{eq:zeta-mass}
\ee
where $\delta m$ vanishes in the limit that sectors 1 and 2 are 
sequestered from each other. 
We note that the ratio $m_\zeta/m_{3/2} = 2$ is truly a SUGRA prediction, 
and measuring this ratio would give valuable insight into the structure 
of spacetime, independent of details of the mechanism of SUSY breaking.

Below we will also consider scenarios in which sectors 1 and 2 both couple 
to the SSM, in which case the sequestered limit is only an approximation. 
While direct interactions between sectors 1 and 2 are induced through 
loops of SSM fields, the resulting $\delta m$ is generally a loop factor 
or more down in size from SSM soft masses~\cite{Cheung:2010mc} and can 
be ignored in most of the parameter regions we will be interested.

\section{Gravitino and Goldstini at Colliders}
\label{sec:collider}

In this section, we will consider the collider phenomenology of LOSP 
decays to gravitinos and goldstini.  For simplicity, we focus on the 
case of two SUSY breaking sectors, with $F_1 > F_2$.  We will be concerned 
with the regime in which $\tilde{m}_1^2 \lesssim \tilde{m}_2^2$, so that 
the SSM fields couple more strongly to the uneaten goldstino $\zeta$ 
than to the gravitino $\tilde{G}$.  The opposite regime has 
phenomenology which is essentially identical to that of standard SUSY.

Consider the limiting case $\tilde{m}_1^2 \ll \tilde{m}_2^2$; extensions 
to more general cases  are straightforward.  In this limit, the partial 
widths of the LOSP into the gravitino and the goldstino take particularly 
simple forms.  As in \Sec{sec:general}, we assume a charged slepton LOSP, 
so
\bea
  \Gamma_{\tilde{\ell} \rightarrow \ell \tilde{G}} 
  &\simeq& \frac{m_{\tilde{\ell}}^5}{16\pi \Feff^2},
\label{eq:gamma-grav}\\
  \Gamma_{\tilde{\ell} \rightarrow \ell \zeta} 
  &\simeq& \frac{m_{\tilde{\ell}}^5}{16\pi F_2^2},
\label{eq:gamma-gold}
\eea
where we have dropped phase space factors for simplicity.
Using the formulas from \Sec{sec:general}, let us now determine the region 
of parameter space in which the BBN bound is satisfied and the 
gravitino (and goldstino) masses can be measured at the LHC.  As discussed 
in Ref.~\cite{Cheung:2010mc}, the decay rate of goldstinos to gravitinos 
is cosmological and therefore irrelevant for our discussions here.

\begin{figure*}[t]
$\Gamma_{\tilde{\ell} \rightarrow 
 \ell \tilde{G}}/\Gamma_{\tilde{\ell} \rightarrow \ell \zeta}$ for $m_{\tilde\ell} = 100~{\rm GeV}$\\
 ~\\
 \includegraphics[scale=0.90]{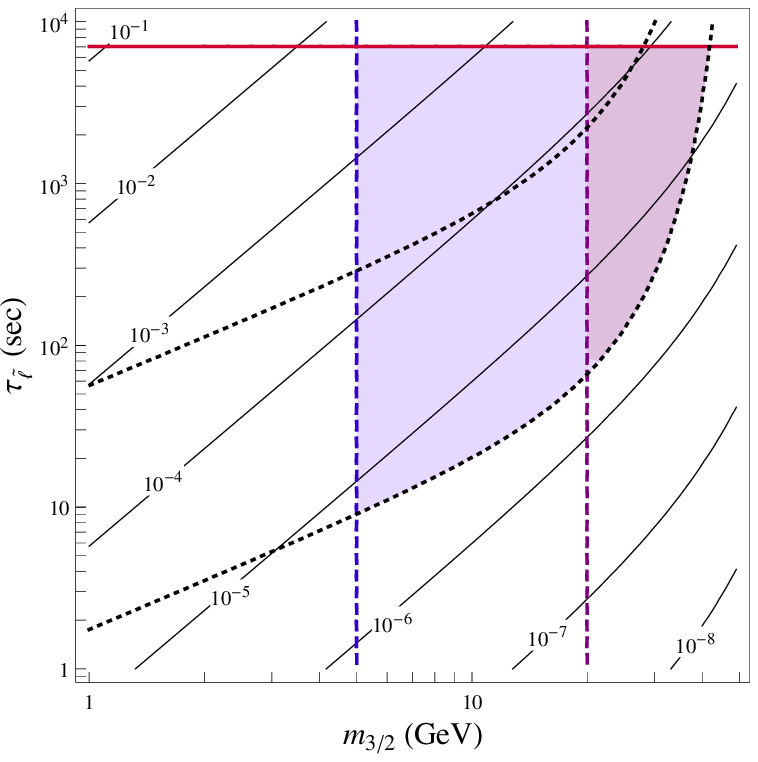} $\qquad$
 \includegraphics[scale=0.90]{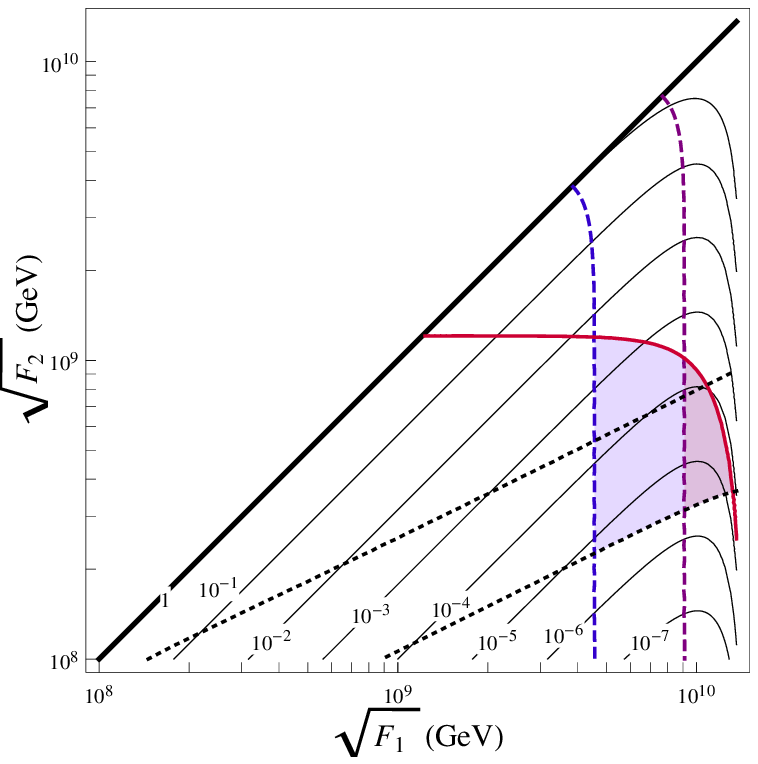}
\caption{Contours of the branching ratio $\Gamma_{\tilde{\ell} \rightarrow 
 \ell \tilde{G}}/\Gamma_{\tilde{\ell} \rightarrow \ell \zeta}$ (labeled, 
 solid black) together with constraints from cosmology and collider 
 physics for $m_{\tilde\ell} = 100~{\rm GeV}$, shown in the 
 $m_{3/2}$-$\tau_{\tilde{\ell}}$ plane (left) and in the 
 $\sqrt{F_1}$-$\sqrt{F_2}$ plane (right).  The BBN bound excludes 
 the parameter regions above the solid red lines, while goldstino 
 overproduction from SSM sfermion decays excludes the regions below 
 the dotted lines (the two dotted lines in each plot correspond to 
 $r \equiv m_{\tilde{Q}}/m_{\tilde{L}} =3$ (lower) and $10$ (upper); see 
 \Sec{subsec:reheating}).  Demanding that the gravitino is heavy enough 
 to be measured at colliders places a lower bound on the gravitino mass 
 depending on experimental resolutions, restricting to the regions right 
 of the vertical dashed lines (blue for $m_{3/2} > 0.05 m_{\tilde{\ell}}$ 
 and purple for $m_{3/2} > 0.2 m_{\tilde{\ell}}$).  The parameter regions 
 consistent with all the constraints are shaded.  To read off analogous 
 bounds on the conventional SUSY setup, simply restrict to the line 
 $F_1 = F_2 \equiv F$.}
\label{fig:m-100GeV}
\end{figure*}
\begin{figure*}[t]
$\Gamma_{\tilde{\ell} \rightarrow 
 \ell \tilde{G}}/\Gamma_{\tilde{\ell} \rightarrow \ell \zeta}$ for $m_{\tilde\ell} = 300~{\rm GeV}$\\
 ~\\
 \includegraphics[scale=0.90]{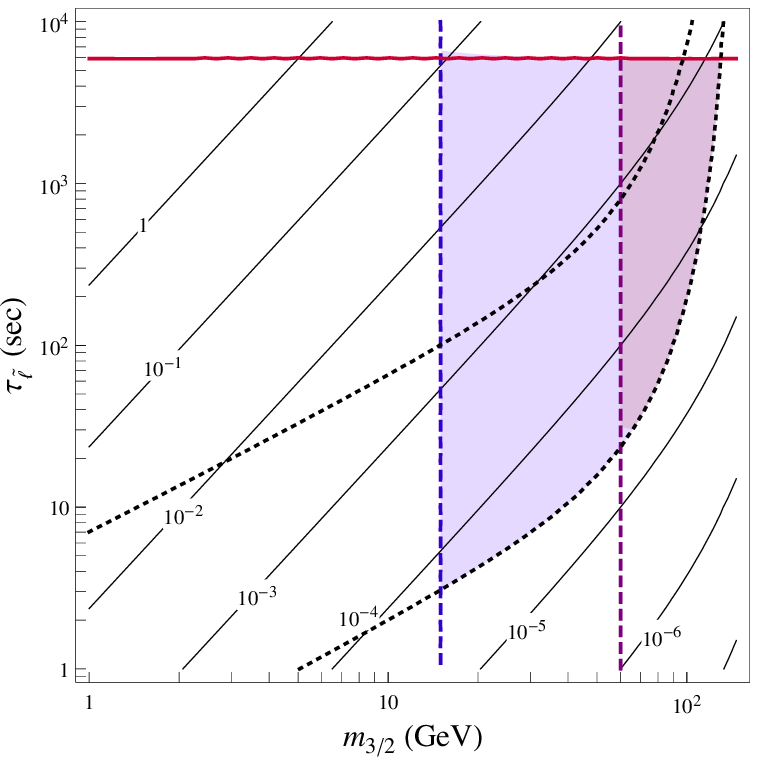} $\qquad$
 \includegraphics[scale=0.90]{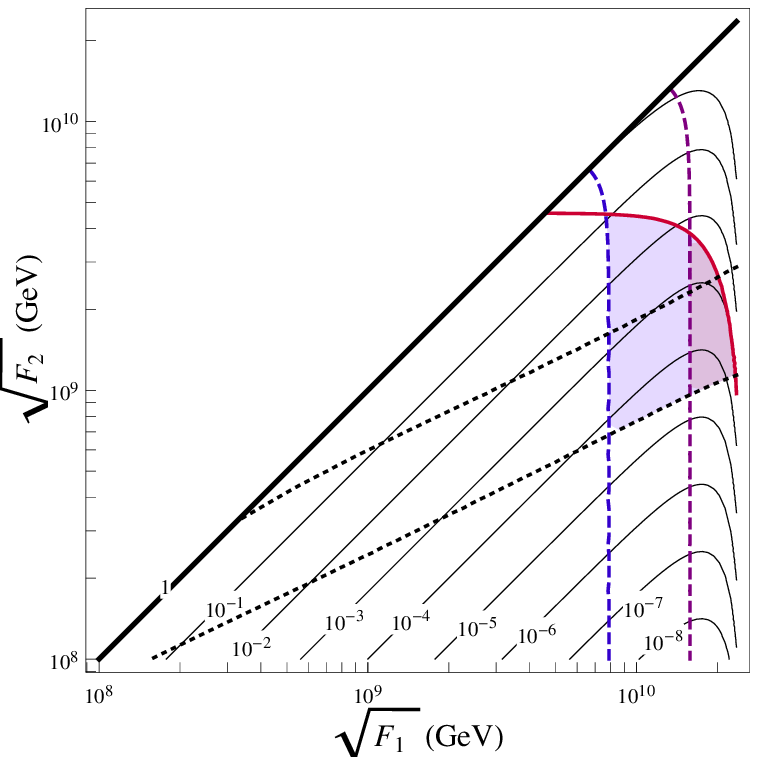}
\caption{The same as \Fig{fig:m-100GeV} but for $m_{\tilde{\ell}} 
 = 300~{\rm GeV}$.}
\label{fig:m-300GeV}
\end{figure*}
The regions of parameter space which satisfy \Eqs{eq:BBN}{eq:mass-obs} 
are shown in \Fig{fig:m-100GeV} (\Fig{fig:m-300GeV}) for $m_{\tilde{\ell}} 
= 100~{\rm GeV}$ ($300~{\rm GeV}$).  The left and right panels depict 
these allowed regions in the $m_{\rm 3/2}$--$\tau_{\tilde{\ell}}$ and 
$F_1$--$F_2$ planes, respectively, where $\tau_{\tilde{\ell}}$ is the 
LOSP lifetime.  In producing these plots, we have included the phase 
space factors and higher order terms in $F_2/F_1$ which are omitted in 
\Eqs{eq:gamma-grav}{eq:gamma-gold}.  In each plot, the region below the 
solid line is allowed by BBN, while the regions right of the vertical, 
dashed lines satisfy \Eq{eq:mass-obs} (with the two lines corresponding 
to $m_{3/2}/m_{\tilde{\ell}} = 0.05$ and $0.2$).  The two dotted lines 
represent the cosmological bound discussed in the next section.  The 
labeled contours denote the branching ratio of $\Gamma_{\tilde{\ell} 
\rightarrow \ell \tilde{G}}/\Gamma_{\tilde{\ell} \rightarrow \ell \zeta}$, 
which must be sufficiently large if we are to be able to see LOSP decays 
to both gravitinos and goldstinos.

A number of important facts are evident from these plots.  In particular, 
we can immediately see the direct conflict between collider signatures and 
BBN in conventional SUSY by considering the plot on the right panel and 
restricting to the diagonal line $F_1 = F_2 \equiv F$.  As expected, 
along this line, there is no region of parameter space in which the 
gravitino is heavy enough to be measured at colliders and also  
simultaneously consistent with BBN constraints.  However, moving down and 
to the right (regions with $F_2 < F_1$), we find that a viable parameter 
space does open up.  Nonetheless, even this region of parameter space is 
limited by cosmological considerations, as we will see in the next section. 
The viable parameter space is thus a finite region in the $F_1$--$F_2$ (and 
$m_{\rm 3/2}$--$\tau_{\tilde{\ell}}$) plane, so that the branching ratio 
$\Gamma_{\tilde{\ell} \rightarrow \ell \tilde{G}}/\Gamma_{\tilde{\ell} 
\rightarrow \ell \zeta}$ has a lower bound, which is of $O(10^{-5})$ 
or so.  This value is not far from the limit of LHC observability, 
given in \Eq{eq:rate-obs}.  It is also interesting that resulting LOSP 
lifetimes, $\tau_{\tilde{\ell}} \approx O(1~\mbox{--}~10^4~{\rm sec})$, 
are within the range in which stopped LOSP decays may be observed in 
the main detector~\cite{Asai:2009ka}.

In order to measure the Planck scale from decays of long-lived charged LOSPs, 
we form a combination of the LOSP mass and lifetime and the mass of the 
invisible LOSP decay product; see \Eq{eq:measure-MPl}.  If the decay 
product is indeed the gravitino, this should reproduce the true Planck 
scale, $\MPl$.  In our case, however, the LOSP decays mainly into $\zeta$, 
so that the measured ``Planck scale'', ``$\MPl$'', will deviate from 
$\MPl$ by
\be
  \mbox{``$\MPl$''}^2 \equiv \frac{m_{\tilde{\ell}}^5} 
    {48\pi \Gamma_{\tilde{\ell} \rightarrow \ell \zeta} m_\zeta^2} 
  = \MPl^2 \frac{\Gamma_{\tilde{\ell} \rightarrow \ell \tilde{G}}}
    {\Gamma_{\tilde{\ell} \rightarrow \ell \zeta}} 
    \frac{m_{\tilde{G}}^2}{m_\zeta^2},
\label{eq:wrong-MPl}
\ee
where again we have dropped phase space factors for the sake of clarity.
Consequently, we expect to measure a value for ``$\MPl$'' which is slightly (one or two orders of magnitude) 
lower than $\MPl$.  Interestingly, the measured value of 
``$\MPl$'' can be used to {\it precisely} fix the branching ratio of the LOSP to the gravitino
\be
  \frac{\Gamma_{\tilde{\ell} \rightarrow \ell \tilde{G}}}
    {\Gamma_{\tilde{\ell} \rightarrow \ell \zeta}} 
  \simeq 4 \left( \frac{\mbox{``$\MPl$''}}{\MPl} \right)^2,
\label{eq:Br-pred}
\ee
where $m_\zeta \simeq 2 m_{3/2}$ has been used.  Thus, by measuring ``$\MPl$'' we know how many stopped LOSPs are necessary to observe the second peak in 
$E_\ell$ which corresponds to the gravitino.  The Planck scale constructed from this second peak should then 
reproduce the value obtained by macroscopic measurements, $\MPl$.

\section{Viable Cosmologies}
\label{sec:cosmology}

It is reasonable to ask to what extent the collider signature discussed 
in the previous section is consistent with cosmology.  For example, if 
the reheating temperature, $T_R$, is smaller than the SSM superparticle 
mass scale, then the only constraints on the masses and couplings of 
gravitinos and goldstinos come from BBN.  If $T_R$ is smaller 
than the LOSP freezeout temperature, then even the constraint from BBN 
disappears.

However, most standard cosmologies require a significantly higher 
reheating temperature, in which case one must evade constraints from the
overproduction of gravitinos and goldstinos as well as the BBN bound. 
We discuss these constraints in \Sec{subsec:reheating}, and present 
a number of consistent cosmological scenarios with high $T_R$ in 
the subsequent subsections.  In each setup, either the goldstino 
or the gravitino could comprise the dark matter of the universe.  Throughout 
this section, we assume the absence of significant entropy production 
below $T_R$, which is indeed the case for most standard cosmologies.

\subsection{Reheating Bounds on Goldstini Couplings}
\label{subsec:reheating}

Avoiding goldstino/gravitino overproduction in
the early universe may provide bounds on $T_R$ and their interactions. 
In the case of the gravitino, this sets a robust upper bound on $T_R$ as 
a function of $m_{3/2}$~\cite{Moroi:1993mb,Rychkov:2007uq}.  In contrast, the bounds from goldstino overproduction depend on 
the goldstino interactions with the SSM fields---unlike the gravitino, the goldstino can have couplings to the SSM which are not universal.

Suppose that sector~2 provides soft mass contributions to all the SSM superparticles. 
In this case, the goldstino couples to SSM fields (almost) universally. 
Since its couplings are larger than those of the gravitino by a factor 
of $\Feff/F_2$, the overproduction bound is correspondingly more 
stringent.  In particular, the standard gravitino overproduction 
bound can be straightforwardly translated into a bound on goldstino overproduction 
via Ref.~\cite{Cheung:2010mc}
\be
  T_R^{\max} \approx 
    10^5~{\rm GeV}\, \Biggl( \frac{10~{\rm GeV}}{m_\zeta} \Biggr) 
    \Biggl( \frac{\sqrt{F_2}}{10^9~{\rm GeV}} \Biggr)^4,
\label{eq:TR-bound}
\ee
for $T_R^{\max}$ larger than the SSM superparticle masses.  As in 
the case of the gravitino, the production of goldstinos in this 
case is dominated at high temperatures by processes involving 
gauge--gaugino--goldstino vertices.  This is because these vertices 
are dimension~5 operators.

As shown in \Sec{subsec:R-cosmo}, however, the gauge--gaugino--goldstino 
interactions can be effectively removed using an $R$ symmetry.  In 
this case, the bound from cosmological goldstino overproduction is far 
milder, since the goldstino couples very weakly to the gauginos.  Instead, 
the leading overproduction bound arises from processes involving the 
scalar--fermion--goldstino couplings, which are dimension~4 interactions. 
Since the strength of these interactions do not grow with temperature, 
the production of goldstinos through decays and scatterings involving SSM 
states is dominated by the infrared.  Thus, the primary constraint 
from goldstino overproduction is a $T_R$-independent bound on the 
scalar--fermion--goldstino couplings.

As discussed in the Appendix, the leading contribution to goldstino 
production through scalar--fermion--goldstino vertices comes from 
superparticle decays.   Since the relevant amplitudes scale with $1/F_2$, 
this sets a lower bound on $m_\zeta/F_2^2 \propto \Feff/F_2^2$.  The 
precise bound depends on the spectrum of superparticles (since the 
goldstino couplings depend on the superparticle masses), and for 
concreteness, we consider $m_{\tilde{Q}} = r m_{\tilde{L}}$ with $r=3$ 
and $10$, where $m_{\tilde{Q},\tilde{L}}$ are the squark and slepton 
masses taken, for simplicity, to be universal at the weak scale. 
These bounds are depicted in \Figs{fig:m-100GeV}{fig:m-300GeV} as 
dotted lines, and given roughly by
\be
  \frac{F_2^2}{\Feff} \simgt 10^{14}~{\rm GeV}^2 
    \left( \frac{m_{\tilde{Q}}}{300~{\rm GeV}} \right)^3.
\label{eq:F2-bound}
\ee

\subsection{The Minimal Goldstini Scenario}
\label{subsec:min-cosmo}

\begin{figure}
  \includegraphics[scale=0.8]{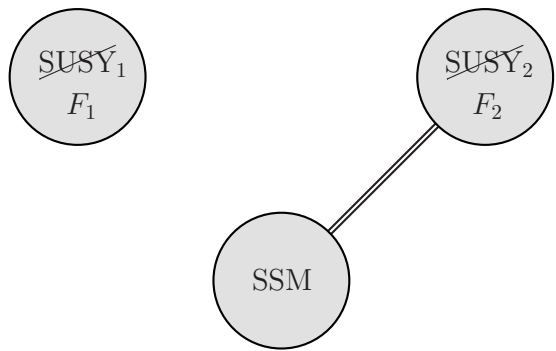}
\caption{Minimal setup in which a standard SUSY breaking scheme 
 (SSM + sector~2) is augmented by an additional sequestered sector 
 which happens to break SUSY at some higher scale (sector~1).}
\label{fig:min-setup}
\end{figure}
In the minimal goldstini scenario, the SSM couples directly to 
sector~2 but not to sector~1, so that $\tilde{m}_1^2$ and $M_1$ both 
vanish (see \Fig{fig:min-setup}).  This corresponds to a setup in which 
a standard SUSY breaking scheme (i.e.\ SSM~$+$~sector~2) is augmented 
by a single sequestered sector which happens to break SUSY at some 
higher scale (i.e.\ sector~1).  Such constructions are expected to 
arise rather naturally from ultraviolet theories.

Since this minimal setup contains gauge--gaugino--goldstino vertices, 
the bound on the reheating temperature from \Eq{eq:TR-bound} applies. 
Consequently, the scalar--fermion--goldstino couplings are important 
only when $T_R^{\max}$ is close to the SSM superparticle masses, 
yielding $O(1)$ corrections to \Eq{eq:TR-bound}.  For low enough $T_R$, 
only the $T_R$-independent bound from \Eq{eq:F2-bound} is relevant, 
as depicted in \Figs{fig:m-100GeV}{fig:m-300GeV}.

When the bound of \Eq{eq:TR-bound} is saturated, $T_R \approx T_R^{\max}$, 
the goldstino comprises all of the dark matter in the universe.  The 
production is dominated by the ultraviolet and is thus sensitive to the 
value of $T_R$.  Comparing \Eq{eq:TR-bound} to the allowed regions 
in  \Figs{fig:m-100GeV}{fig:m-300GeV}, we find that $T_R$ as high as 
$\approx 10^7~{\rm GeV}$ (corresponding to $m_\zeta \simeq \mbox{few} 
\times 10~{\rm GeV}$, $F_2 \simeq \mbox{few} \times 10^9~{\rm GeV}$) 
can be consistent with the BBN bound and the collider measurement 
of $\MPl$.  Such a high reheating temperature allows for high 
temperature mechanisms for baryogenesis that would otherwise not work with 
$T_R \lesssim m_{\rm LOSP} / 20$, although 
it is still too low for thermal leptogenesis.  Note that the gravitino 
abundance is small, $\Omega_{3/2} \approx (F_2/\Feff)^2 \Omega_\zeta$, 
and that the goldstino energy density coming from late decays of the 
LOSP after LOSP freezeout is also typically subdominant.

\subsection{SUSY Breaking with $R$ Symmetries}
\label{subsec:R-cosmo}

Copious production of goldstinos at very high temperatures is not 
inevitable.  In particular, since the bound on the reheating temperature 
\Eq{eq:TR-bound} is wholly determined by the goldstino couplings to the 
gauginos, it can be evaded by imposing an $R$ symmetry.  This allows 
for an alternative cosmological scenario with non-thermal gravitino 
dark matter.

\begin{figure}
  \includegraphics[scale=0.8]{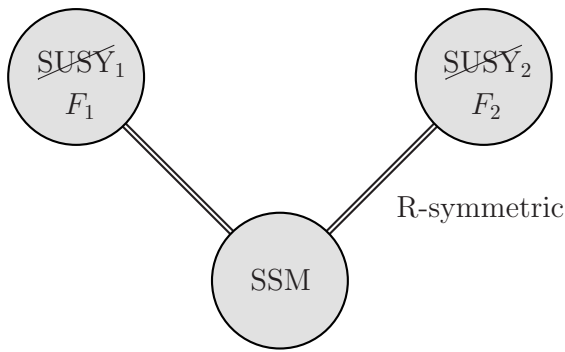}
\caption{Setup in which both sector~1 and sector~2 couple directly to 
 the SSM.  By construction, couplings between sector~2 and the SSM are 
 $R$ symmetric, so gaugino masses arise solely from sector~1.  For the 
 same reason, the goldstino has suppressed couplings to gauginos.}
\label{fig:R-setup}
\end{figure}
Consider the setup depicted in \Fig{fig:R-setup}, where sector~2 preserves 
an $R$ symmetry.  In this case, the sfermion masses receive a contribution 
from sector~2, $\tilde{m}_2^2 \neq 0$, but not the gaugino masses, 
$M_2 = 0$.  Sector~1, which does not preserve an $R$ symmetry, generates 
both $\tilde{m}_1^2$ and $M_1$.  For simplicity, we consider that the 
resulting sfermion and gaugino masses are of the same order, which 
can easily happen if $\tilde{m}_2^2$ is not much larger than $M_1^2$. 
Our analysis below assumes that the dominant contribution to the sfermion 
masses comes from sector~2, although the existence of a comparable 
contribution from sector~1 does not change our essential conclusions.

In this $R$-symmetric setup, the bound from goldstino overproduction 
is quite mild because the goldstino couples very weakly to the gauginos. 
The only relevant interactions are the scalar--fermion--goldstino 
couplings, so we need only consider the $T_R$-independent bound from 
\Eq{eq:F2-bound} which are shown in \Figs{fig:m-100GeV}{fig:m-300GeV}. 
It is interesting that this cosmological bound leads to a lower limit 
on the branching fraction $\Gamma_{\tilde{\ell} \rightarrow \ell 
\tilde{G}}/\Gamma_{\tilde{\ell} \rightarrow \ell \zeta} \simgt O(10^{-5})$, 
which favors the possibility of observing both $\tilde{G}$ and $\zeta$ 
at the LHC.

Since the coupling strength of the gauginos to the goldstino is 
a factor of $F_2/F_1$ weaker than to the gravitino in the present 
setup, the constraint from goldstino overproduction through these 
couplings is weaker than that from gravitino overproduction.  From 
\Figs{fig:m-100GeV}{fig:m-300GeV}, parameter regions we are interested 
in are roughly $F_2 \sim 10^9~{\rm GeV}$ and $\Feff \approx F_1 \sim 
10^{10}~{\rm GeV}$.  The bound on $T_R$ from gravitino overproduction in 
these parameter regions is rather weak~\cite{Moroi:1993mb,Rychkov:2007uq}
\be
  T_R^{\max} \approx O(10^8~\mbox{--}~10^{10}~{\rm GeV}),
\label{eq:TR-high}
\ee
so that it can even be compatible with thermal leptogenesis, which typically 
requires $T_R \simgt 10^9~{\rm GeV}$.  If the bound of \Eq{eq:TR-high} is 
saturated, we have gravitino dark matter.

Note that in conventional SUSY breaking scenarios, gravitino dark matter 
with a high reheating temperature such as in \Eq{eq:TR-high} is not 
possible, due to stringent constraints from BBN.  In our case, however, 
the LOSP decays to the goldstino faster than to the gravitino as long as 
$\tilde{m}_2^2/F_2$ is sufficiently large.  This allows us to evade the 
BBN bound consistently with gravitino dark matter and thermal leptogenesis.

\subsection{Late Decay Case}
\label{subsec:late-decay}

So far, we have assumed that the relic density of goldstinos arising 
from late LOSP decays is small.  This is true in most of the natural parameter 
regions, but in certain corners of parameter space, goldstinos from late 
LOSP decays may saturate the observed dark matter abundance.

Suppose that the slepton freezeout abundance is completely controlled 
by annihilation into gauge bosons (which will be the case if the 
neutralinos and heavy Higgs bosons are sufficiently heavier than the 
slepton).  In this case, the yield of the (mostly right-handed) slepton 
before its decay is given by $Y_{\tilde{\ell}} \simeq 2 \times 10^{-13} 
(m_{\tilde{\ell}}/100~{\rm GeV})$~\cite{Berger:2008ti}.  This leads 
to the goldstino relic abundance
\be
  \Omega_\zeta \simeq 0.2
    \left( \frac{m_\zeta}{200~{\rm GeV}} \right) 
    \left( \frac{m_{\tilde{\ell}}}{1~{\rm TeV}} \right),
\ee
so that if the slepton is very heavy, dark matter goldstinos may mostly 
come from late slepton decays.  Such heavy sleptons, however, may be problematic for LHC 
measurements.

\section{Other LOSPs}
\label{sec:others}

In the previous discussion, we have considered the case where the LOSP 
is a (mostly right-handed) charged slepton.  In this section, we briefly 
discuss other possibilities.

To achieve the signatures discussed in this paper, there must be 
a quasi-stable charged state which is stopped either in the main 
detector or in a stopper detector.  This immediately eliminates 
the possibility of a (mostly) bino LOSP.  Similarly, wino, Higgsino 
or left-handed slepton LOSPs typically do not lead to the relevant 
signatures, since a mass splitting between the charged and neutral 
components (induced by radiative corrections, tree-level mixings, 
or the $D$-term effect) are so large that the charged component 
decays with lifetime shorter than $\approx 10^{-6}~{\rm sec}$ (see 
however~\cite{Kribs:2008hq}).  With the constraint from overproduction 
in \Sec{subsec:reheating}, the branching fraction of a charged 
component to the goldstino is then tiny, $\simlt 10^{-6}$.

This leaves only the possibility of a gluino or squark LOSP.  In 
the early universe 
these states freeze out at a temperature of $O(0.1 m_{\rm LOSP})$, with an abundance determined by perturbative 
strong interaction processes, $Y_{\rm LOSP,pert} \approx 10^{-13} 
(m_{\rm LOSP}/1~{\rm TeV})$.  For the gluino $\tilde{g}$, this abundance 
will be reduced significantly by nonperturbative annihilations occurring 
after the QCD phase transition, $Y_{\tilde{g}} \approx 10^{-20} 
(m_{\tilde{g}}/1~{\rm TeV})^{1/2}$~\cite{Kang:2006yd}.  On the other 
hand, for squarks $\tilde{q}$, nonperturbative processes lead to a 
significant fraction of $\tilde{q}\tilde{q}\tilde{q}$ bound states, 
which are not subject to enhanced annihilations.  Therefore, the 
squark abundance may not be much reduced from the perturbative 
value, $Y_{\tilde{q}} \approx O(10^{-14}~\mbox{--}~10^{-13}) 
(m_{\tilde{q}}/1~{\rm TeV})$.

With the relic abundance given above, the gluino LOSP does not suffer 
from the BBN constraint.%
\footnote{This implies that if the gluino is the LOSP, the collider 
 measurement of $\MPl$ can be consistent with the BBN bound even in 
 the conventional SUSY framework.  The measurement of gluino decays 
 will be discussed below.}
On the other hand, squark LOSPs are subject to the BBN constraint coming 
from hadronic energy injections; conservatively it is $\tau_{\tilde{q}} 
\simlt 100~{\rm sec}$~\cite{Kawasaki:2004qu}.  For a fixed LOSP 
mass, the constraint from goldstino overproduction can be weaker for 
gluino/squark LOSPs than for slepton LOSPs, since the masses of colored 
superparticles, which mainly control the goldstino abundance, can be 
smaller.  A conservative constraint is given by \Eq{eq:F2-bound} with 
$m_{\tilde{Q}}$ replaced by $m_{\rm LOSP}$, which corresponds to taking 
$r \simeq 1$.

Gluino/squark LOSPs can be produced at the LHC either directly or 
through decays of heavier superparticles.  After being produced, they 
hadronize by picking up a gluon $g$ or up/down quarks $q = u,d$.  For 
the gluino, the relevant bound states are $\tilde{g}g$, $\tilde{g}\bar{q}q$, 
and $\tilde{g}qqq$.  While the precise spectrum of bound states is not obvious, a fraction of gluinos is stopped in the detector under reasonable 
assumptions~\cite{Arvanitaki:2005nq}, allowing for gluino decay measurements 
(assuming that tracks can be reconstructed despite charge oscillation). 
The mass of the gluino can also be measured using charged gluino bound 
states traversing the muon system.  The measurement of the Planck scale 
will thus be feasible for the gluino LOSP.  The situation for squark 
LOSPs is similar, where the relevant bound states are $\tilde{q}\bar{q}$ 
and $\tilde{q}qq$.

The visible decay products of gluino/squark LOSPs are jets, with an 
energy resolution expected to be $\varDelta_E \approx O(1\%)$.  Therefore, 
to be able to perform the measurements discussed in this paper, the 
masses of the gravitino and goldstino must be larger than $\approx 
O(0.1) m_{\rm LOSP}$; see \Eq{eq:m32-error}.

In summary, the parameter regions in which the
goldstino/gravitino collider signals are obtained consistently with high reheating 
temperatures (i.e.\  satisfying both the BBN and overproduction 
constraints) are
\be
  \tau_{\tilde{g}} \simgt \tau_{\rm min},
\qquad
  m_{3/2} \simgt O(0.1) m_{\rm LOSP},
\ee
for a gluino LOSP, and
\be
  \tau_{\rm min} \simlt \tau_{\tilde{q}} \simlt 100~{\rm sec},
\qquad
  m_{3/2} \simgt O(0.1) m_{\rm LOSP},
\ee
for a squark LOSP.  Here,
\be
  \tau_{\rm min} = 0.2~{\rm sec} 
    \left( \frac{300~{\rm GeV}}{m_{\rm LOSP}} \right) 
    \left( \frac{m_{3/2}/m_{\rm LOSP}}{0.1} \right),
\ee
is obtained by translating \Eq{eq:F2-bound} into a bound on LOSP lifetimes 
ignoring the phase space factor, which, however, would become important 
when $m_\zeta \approx m_{\rm LOSP}$.

\section{Conclusions}
\label{sec:concl}

The LHC may offer an unprecedented opportunity to probe the fundamental 
structure of spacetime at colliders.  In particular, if the LOSP is 
charged, then precision measurements of its decays to the gravitino 
could provide a genuine collider measurement of $\MPl$ and a dramatic 
confirmation of SUGRA.  Unfortunately, this decay process is directly 
constrained by BBN in the early universe.  Thus, there must be some 
modification of the conventional SUSY framework to allow for high 
reheat temperatures $T_R \simgt {\rm TeV}$ to be consistent with 
collider probes of SUGRA.

In this paper, we have shown that the goldstini framework introduced 
in Ref.~\cite{Cheung:2010mc} provides precisely such a modification. 
Multiple sources of SUSY breaking yield a corresponding multiplicity 
of goldstini which can easily couple more strongly to the SSM than the 
gravitino.  Thus, the LOSP decays to goldstini fast enough to avoid 
the BBN bound, while the gravitino mass can still be measured in 
colliders via the LOSP decay to the gravitino.  In fact, the regions 
in parameter space where this occurs are favored by cosmology.

Intriguingly, within this setup colliders will first measure the LOSP 
decay to the goldstino.  Initially, this will almost certainly be 
interpreted as a LOSP decay to a gravitino, which will in turn result 
in a mismeasurement of $\MPl$ (one or two orders of magnitude below 
the value obtained from long-distance gravity).  As we have shown, 
the degree of the discrepancy actually fixes the LOSP branching ratio 
into the gravitino, and hence the amount of integrated luminosity 
needed to discover the gravitino.  Once this target luminosity is 
reached, our framework can be tested unambiguously.  In particular, one 
may measure the masses of both the gravitino and goldstino, and if these 
satisfy $m_\zeta = 2 m_{3/2}$ as predicted in Ref.~\cite{Cheung:2010mc}, 
then this would provide a smoking gun signature of the goldstini setup. 
Specifically, we would learn not only that SUGRA is a symmetry of 
nature, but also that SUSY is broken multiple times and that sequestering 
is a real phenomenon.  This would in turn suggest the existence of 
compact extra dimensions in which sequestering naturally emerges.

The scenarios described here are consistent with standard cosmology 
with high reheating temperatures.  In particular, if the sector giving 
the goldstino preserves an $R$ symmetry, then the bound from goldstino 
overproduction does not lead to an extra constraint on $T_R$ beyond 
that from gravitino overproduction.  This allows for thermal leptogenesis 
with LSP (gravitino) dark matter, which is not possible in the standard 
SUSY framework with $R$-parity.

\paragraph*{Note Added:} Related work~\cite{DeSimone:2010tr} discussing a similar mechanism for 
evading BBN constraints appeared concurrently with this paper.

\begin{acknowledgments}
The work of C.C., J.M., and Y.N. was supported in part by the Director, Office 
of Science, Office of High Energy and Nuclear Physics, of the US Department 
of Energy under Contract DE-AC02-05CH11231, and in part by the National 
Science Foundation under grants PHY-0555661 and PHY-0855653.  J.T. is 
supported by the U.S. Department of Energy under cooperative research 
agreement DE-FG0205ER41360.
\end{acknowledgments}

\appendix

\section{Infrared-Dominated Goldstino Production}
\label{app:ir-goldstini}

The late-time goldstino yield can be computed with a standard Boltzmann 
equation calculation.  The yield is defined as $Y_{\zeta} \equiv
n_{\zeta}/s$, where $n_{\zeta}$ is the goldstino number density and $s$ is
the total entropy density, and is constant once goldstino production is
completed.  There are three potentially relevant goldstino production 
mechanisms: superparticle decays and $2 \to 2$ scattering processes 
in the early thermal bath, and late decays of relic LOSPs after LOSP 
freezeout.  If the goldstino does not couple to gauge multiplets, 
as occurs in the $R$-symmetric setup described in \Sec{subsec:R-cosmo}, 
then the goldstino production is dominated by scalar decays at 
$T \sim \tilde{m}$, and is insensitive to $T_R$.  This is in contrast 
to the standard gravitino production calculation \cite{Moroi:1993mb}, 
where the goldstino abundance grows linearly with $T_R$.

Here, we briefly describe the calculation of the contribution 
from superparticle decays in this scenario.  The contribution 
from $2 \to 2$ scattering can be calculated in a similar (but more 
involved) manner, but we find it to be subdominant and omit it from 
our analysis.  The contribution from late LOSP decays can be taken 
directly from the LOSP freezeout abundance used to determine 
the BBN bound~\cite{Kawasaki:2008qe}.  For a slepton LOSP, for 
example, $Y_\zeta^\mathrm{(LOSP-decay)} = 7 \times 10^{-14} 
(m_{\tilde{\ell}}/100~{\rm GeV})$, which is not significant unless 
$m_{\tilde{\ell}} \simgt 700~{\rm GeV}$.

The goldstino yield from decays in the thermal bath is found by 
solving the Boltzmann equation:
\be
  \dot{n}_\zeta + 3 H n_\zeta = \sum_{i} n_i 
    \bigg\langle \frac{1}{\gamma} \bigg\rangle_i \Gamma_{i\to\zeta},
\ee
where dots indicate derivatives with respect to time, the sum is 
over unstable species, $n_i$ and $\Gamma_{i\to\zeta}$ are their 
number densities and decay rates to goldstinos, and $\langle 1/\gamma 
\rangle$ is the thermally averaged relativistic time-dilation factor 
to account for out-of-rest-frame decay rates.  Using the fact that 
the entropy per comoving volume is constant, we have the relations
\be
  \dot{Y}_\zeta = \frac{1}{s} \left( \dot{n}_\zeta + 3 H n_\zeta \right),
\ee
and
\be
  \frac{dt}{dT} = -\frac{1}{H T} 
    \left( 1 + \frac{1}{3} \frac{d\log{g_S(T)}}{d\log{T}} \right),
\ee
where $T$ is temperature and $g_S(T)$ is the effective number of 
relativistic species.  The goldstino yield is thus
\be
  Y_\zeta = \int_{T_R}^0 dT \frac{dt}{dT} \dot{Y}_\zeta.
\ee

For concreteness, we assume the simple spectrum $m = m_{\rm LOSP}$ 
for non-colored particles and $m =r m_{\rm LOSP}$ for colored particles, 
with $r$ a free parameter.  Squark decays dominate the production 
process, with a decay width
\be
  \Gamma_{\tilde{Q} \to Q\zeta} \simeq 
    \frac{1}{16\pi} \frac{m_{\tilde{Q}}^5}{F_2^2}.
\ee
Parametrically, for decays
\be
  \dot{Y}_\zeta^\mathrm{decay} \simeq 
    \frac{m_{\tilde{Q}}^5}{F_2^2} \theta(T-m_{\tilde{Q}}),
\qquad
  \frac{dt}{dT} \sim -\frac{\MPl}{T^3}.
\ee
Solving the Boltzmann equations numerically, keeping the full 
temperature dependence, we find:
\be
  Y_\zeta^\mathrm{decay} \approx 
    0.0013 \frac{\MPl r^3 m_{\rm LOSP}^3}{F_2^2}.
\ee
Here, we show only the leading order dependence on $F_1$, $F_2$ and $r$, 
but we keep the full dependence in \Figs{fig:m-100GeV}{fig:m-300GeV}. 
The goldstino overabundance bound is set by requiring $m_\zeta Y_\zeta 
< 3.8 \times 10^{-10}~{\rm GeV}$~\cite{Amsler:2008zzb}, so that the 
goldstino abundance is not in conflict with the observed dark matter 
density.

\end{document}